\documentclass[preprintnumbers,amsmath,amssymb,letterpaper,nofootinbib,notitlepage]{revtex4}
\usepackage{graphicx}% Include figure files
\usepackage{enumerate}% For non-numerical items
\usepackage{bm}% bold math
\usepackage{slashed}

\usepackage{amssymb}
\usepackage{amsmath}
\usepackage{epsfig}
\newcommand{\be}{\begin{equation}}
\newcommand{\ee}{\end{equation}}
\newcommand\beq{\begin{eqnarray}}
\newcommand\eeq{\end{eqnarray}}

\renewcommand{\eqref}[1]{(\ref{#1})}

\newcommand{\mybar}[1]%
        {\kern 0.6pt\overline{\kern -0.6pt#1\kern -0.6pt}\kern 0.6pt}
\newcommand{\ccny}{
	Department of Physics,
	The City College of New York,
	New York, NY 10031, USA
	}
\newcommand{\cuny}{
	Graduate School and University Center,
	The City University of New York,
	New York, NY 10016, USA
	}

\newcommand{\jlabcomp}{
	Scientific Computing Group,
	Thomas Jefferson National Accelerator Facility,
	Newport News, VA 23606, USA
	}

\newcommand{\julich}{
	 Institut f\"{u}r Kernphysik and Institute for Advanced Simulation,
	 Forschungszentrum J\"{u}lich, 54245 J\"{u}lich Germany
 }
\newcommand{\lblnsd}{
	Nuclear Science Division
	Lawrence Berkeley National Laboratory,
	Berkeley, CA 94720, USA
	}
\newcommand{\lblnersc}{
	NERSC
	Lawrence Berkeley National Laboratory,
	Berkeley, CA 94720, USA
	}
\newcommand{\liverpool}{
	Theoretical Physics Division, Department of Mathematical Sciences, University of Liverpool,
	Liverpool L69 3BX, UK
    }
\newcommand{\llnl}{
	Physics Division,
	Lawrence Livermore National Laboratory,
	Livermore, CA 94550, USA
	}
\newcommand{\nvidia}{
    NVIDIA Corporation,
    2701 San Tomas Expressway, Santa Clara, CA 95050, USA
    }
\newcommand{\rbrc}{
	RIKEN BNL Research Center,
	Brookhaven National Laboratory,
	Upton, NY 11973, USA
	}
\newcommand{\ucb}{
	Department of Physics,
	University of California,
	Berkeley, CA 94720, USA
	}
\newcommand{\unc}{
	Department of Physics and Astronomy,
	University of North Carolina,
	Chapel Hill, NC 27516-3255, USA
	}
\newcommand{\wm}{
	Department of Physics,
	The College of William \& Mary,
	Williamsburg, VA 23187, USA
	}
\newcommand{\intuw}{
Institute for Nuclear Theory, University of Washington, 
Seattle, WA 98195, USA}
\newcommand{\glas}{
University of Glasgow, Glasgow G12 8QQ, UK}
\newcommand{\ithems}{
Interdisciplinary Theoretical and Mathematical Sciences Program (iTHEMS),
RIKEN 2-1 Hirosawa,
Wako, Saitama 351-0198, Japan
}

\begin{document}
\title{Symmetries and Interactions from Lattice QCD}

\author{A.~Nicholson}
\email{annichol@email.unc.edu}
\affiliation{\unc}
\affiliation{\ucb}

\author{E.~Berkowitz}
\affiliation{\julich}

\author{H.~Monge-Camacho}
\affiliation{\unc}
\affiliation{\wm}
\affiliation{\lblnsd}

\author{D.~Brantley}
\affiliation{\wm}
\affiliation{\llnl}
\affiliation{\lblnsd}

\author{N.~Garron}
\affiliation{\liverpool}

\author{C.C.~Chang}
\affiliation{\ithems}
\affiliation{\ucb}
\affiliation{\lblnsd}

\author{E.~Rinaldi}
\affiliation{\rbrc}
\affiliation{\lblnsd}

\author{C.~Monahan}
\affiliation{\intuw}

\author{C.~Bouchard}
\affiliation{\glas}

\author{M.A.~Clark}
\affiliation{\nvidia}

\author{B.~Jo\'{o}}
\affiliation{\jlabcomp}

\author{T.~Kurth}
\affiliation{\lblnersc}

\author{B.C.~Tiburzi}
\affiliation{\ccny}
\affiliation{\cuny}

\author{P.~Vranas}
\affiliation{\llnl}
\affiliation{\lblnsd}

\author{A.~Walker-Loud}
\affiliation{\lblnsd}
\affiliation{\llnl}
\affiliation{\ucb}

%\collaboration{CalLat}
%\noaffiliation

\date{\today}

\preprint{LLNL-CONF-764382, RBRC-1296, RIKEN-iTHEMS-Report-18, INT-PUB-18-063}%, KITP-XXXX}
% ------------------------------------------------------------------

% ------------------------------------------------------------------
\begin{abstract}
Precision experimental tests of the Standard Model of particle physics (SM) are one of our best hopes for discovering what new physics lies beyond the SM (BSM). Key in the search for new physics is the connection between theory and experiment. Forging this connection for searches involving low-energy hadronic or nuclear environments requires the use of a non-perturbative theoretical tool, lattice QCD. We present two recent lattice QCD calculations by the CalLat collaboration relevant for new physics searches: the nucleon axial coupling, $g_A$, whose precise value as predicted by the SM could help point to new physics contributions to the so-called ``neutron lifetime puzzle", and hadronic matrix elements of short-ranged operators relevant for neutrinoless double beta decay searches.
\vfill
\noindent Plenary talk presented at the Thirteenth Conference on the Intersections of Particle and Nuclear Physics, 
Palm Springs, CA  May 28 - June 3, 2018
\end{abstract}

\maketitle
% ------------------------------------------------------------------

%%%%%%%%%%%%%%%%%
%%%%%%%%%%%%%%%%%
\section{Introduction}
%%%%%%%%%%%%%%%%%
%%%%%%%%%%%%%%%%%

The Standard Model (SM) of particle physics stands as a crowning achievement of physicists during the 20th century. However, it is well accepted that the SM must be incomplete, as it is unable to answer such outstanding questions as:
\begin{itemize}
\item
What is the origin of neutrino masses?
\item
What is the nature of dark matter?
\item
Why does the universe contain an excess of matter over antimatter?
\end{itemize}
To provide some answers, we must develop new theories which extend beyond the SM (BSM), and over the years theorists have proposed a plethora of possibilities. Progress in understanding BSM physics relies heavily on experimental efforts, which can rule out or confirm theoretical predictions, as well as provide new unexplained data which may act as clues toward determining the correct theory. 

The ability to faithfully connect theory to experiment is crucial for these efforts. For low energy experiments utilizing nuclear or hadronic environments, forging this connection requires the solution of the theory behind the strong nuclear force, Quantum Chromodynamics (QCD). Unfortunately, attempts to directly determine analytical solutions of QCD at low energies are thwarted by the large value of the strong coupling constant in this regime, rendering perturbation theory unreliable. Currently our only tool for tackling this enterprise which can produce accurate results with fully controlled and quantifiable systematics is a numerical technique known as Lattice QCD (LQCD). LQCD provides non-perturbative solutions to the QCD path integral through the use of a lattice regulator, in which spacetime is discretized. Physical results are obtained through removal of this regulator, by taking the continuum limit using well-understood prescriptions. In order to fit the resulting infinite-dimensional integral into a finite-sized computer, only a finite volume of spacetime may be studied, and frequently heavier-than-physical quark masses (and therefore, pion masses) are used. Thus, final results are obtained through the continuum, infinite volume, and physical pion mass limits. 

LQCD contributions to experimental searches for BSM physics may be roughly divided into two categories. The first type of calculation involves producing precision results for quantities which may be equally precisely determined experimentally. By comparing theoretical expectations from the SM to experimental findings, one may look for discrepancies as signs of new physics. An example is the muon anomalous magnetic moment, which currently shows an approximately $3.5\sigma$ discrepancy between theory and experiment~\cite{Tanabashi:2018oca} (for a review, see Ref.~\cite{Blum:2013xva}). Also compelling are quantities for which an experimental ``puzzle" exists: two different types of experiments produce results that are in tension with one another. Examples are the ``proton radius puzzle" \cite{Carlson:2015jba,Hill:2017wzi,doi:10.1146/annurev-nucl-102212-170627,Jentschura:2010ha}, involving experiments utilizing different leptonic probes, and the ``neutron lifetime puzzle" \cite{Tanabashi:2018oca, Czarnecki:2018okw}, arising between measurements of trapped neutrons versus beams of neutrons. Reducing the uncertainty on the theoretical determination of the latter requires a precise theoretical determination of the value of the nucleon axial coupling, $g_A$. In Sec.~\ref{sec:gA}, we will discuss a new calculation of $g_A$ to $1\%$ precision which puts a strong theoretical constraint on right-handed BSM operators, and paves the way toward sub-percent calculations capable of resolving the discrepancy between lifetime experiments~\cite{Chang:2018uxx,Berkowitz:2018gqe}.

A complimentary experimental program searching for BSM physics involves the use of nuclear or hadronic probes looking for extraordinarily rare processes. Such processes are forbidden by the symmetries of the SM, and would therefore point to new physics. Examples are dark matter searches, which use nuclear detectors to look for interactions with non-SM particles. These interactions occur through the so-called nucleon sigma terms, which may be determined through LQCD calculations~\cite{FLAG:2019}. LQCD calculations of permanent hadronic electric dipole moments (for recent reviews, see Refs.~\cite{Yoon:2017tag,Syritsyn:2018mon}) arising from both SM and potential BSM contributions will aid in searches for CP violation, a necessary ingredient in the production of a matter/antimatter asymmetry in the universe. Also necessary for this asymmetry is baryon number violation. Experimentalists are searching for the proposed neutrinoless double beta decay ($0\nu\beta\beta$), which, if observed, would elucidate the origins of this symmetry violation in the early universe. LQCD calculations of these decays in few hadron systems are a necessary step for determining the potential sensitivity of current and planned experiments, as well as connecting any signals to the underlying BSM theory at play~\cite{Tiburzi:2017iux,Shanahan:2017bgi,Detmold:2018zan,Feng:2018pdq}. In Sec.~\ref{sec:0nubb} we will discuss a calculation of hadronic matrix elements for short-ranged decay operators, arising from new heavy physics, which may lead to a greatly modified expectation for the potential discovery region of $0\nu\beta\beta$ experiments~\cite{Nicholson:2018mwc}. 

\section{\label{sec:gA}Axial coupling of the nucleon}

The axial coupling of the nucleon, $g_A$, is a central quantity in nuclear physics. The axial coupling determines the coupling of the weak axial current to the nucleon, and is therefore fundamental in the calculation of the neutron weak decay, and by extension, nuclear beta decay. According to the SM, the free neutron lifetime, $\tau_n$, and axial coupling are related by~\cite{Czarnecki:2018okw},
\beq
\left|V_{ud}\right|^2 \tau_n \left(1+3g_A^2\right) = 4908.6(1.9) \mathrm{~s} \ ,
\eeq
where $\left|V_{ud}\right|$ is the CKM mixing element\footnote{A new analysis of nuclear structure effects by the authors of Ref.~\cite{Seng:2018qru} has determined a slightly lower value for the numerical constant.}. The axial coupling is equally important for other weak processes involving nucleons, such as the proton-proton fusion reaction ($pp\to d\gamma$) which powers our sun. The amount of hydrogen versus helium in the universe following the Big Bang also depends very sensitively upon $g_A$, thus its precise value heavily influences Big Bang nucleosynthesis. 

Finally, through the Goldberger-Treiman relation,
\beq
\label{eq:GT}
g_A m_n = F_{\pi} g_{\pi nn} \ ,
\eeq
where $m_n$ is the mass of the nucleon and $F_{\pi}$ is the pion decay constant, the pion-nucleon coupling, $g_{\pi nn}$ is also related to $g_A$. This coupling controls pion-nucleon scattering cross sections, the long-range strong nuclear force, and an array of related nuclear physics observables. Because it determines the interactions between pions and nucleons, it is ubiquitous in Chiral Perturbation Theory calculations for nucleons, and is therefore crucial for determining the quark mass dependence of quantities involving nucleons.

The axial coupling has been very precisely measured experimentally. For example, a recent measurement by the UCNA group of the polarized neutron spin-electron correlation coefficient,
\beq
A_0(\lambda) = \frac{-2\left(\lambda^2-|\lambda|\right)}{1+3\lambda^2} \ ,
\eeq
where $\lambda = g_A/g_V$, yielded $\lambda = -1.2783(22)$ \cite{Brown:2017mhw}, while an even more recent measurement by the Perkeo group yielded $\lambda = -1.27641(45)(33)$~\cite{Markisch:2018ndu}. Through chiral symmetry, the SM predicts $g_V = 1 + \mathcal{O}\left((m_u-m_d)^2\right)$, therefore, the quantity $\lambda$ directly predicts $g_A$, barring BSM effects. The current average given by the Particle Data Group is $g_A = 1.2724(23)$~\cite{Tanabashi:2018oca}. There has, however, been a notable trend over time of increasing values of experimental measurements of $g_A$, leading to a $5\sigma$ difference between average values obtained using pre- and post-2002 measurements \cite{Czarnecki:2018okw}. 

While this discrepancy may be attributed to the potentially increased difficulty in estimating systematics for pre-2002 experiments, perhaps more troubling is a nearly $4\sigma$ discrepancy between independent experimental measurements of the related quantity $\tau_n$ utilizing different experimental strategies \cite{Serebrov:2017bzo}. In one type of experiment, ultracold neutrons are contained via magnetic or gravitational traps, and the number that remain after a period of time is tracked. In another, beams of neutrons are used, and the number of protons that are produced is recorded. Thus, beam experiments count only the number of decays of neutrons into protons, as predicted by the SM, while trap experiments can also account for potential non-SM decays. The smaller lifetime measured by trap experiments could potentially be pointing to new physics in these decays, for example, neutron decays into dark matter~\cite{Fornal:2018eol,Tang:2018eln,Serebrov:2018mva}.

A direct calculation of $g_A$ from the SM could help clarify these experimental pictures, with a $0.5\%$ calculation potentially showing favor for one type of experiment over the other, and a $0.1\%$ calculation showing $5\sigma$ selective resolution. Lattice QCD (combined with QED) is the only known method for carrying out such a calculation. Theoretical motivation to calculate $g_A$ using lattice QCD extends far beyond these goals. There are related quantities for which current experimental measurements are insufficiently determined, such as axial form factors and in-medium corrections to $g_A$. Axial form factors are a crucial limiting factor in the precision of long-baseline neutrino experiments, such as the upcoming DUNE experiment at Fermilab, which will probe sources of CP violation in the neutrino sector. In-medium effects on axial currents may prove important for understanding $0\nu\beta\beta$ experiments, which will be discussed in the next Section. Lattice calculations of these quantities have been performed at unphysical pion mass \cite{Savage:2016kon,Savage:2016mlr,Chang:2017eiq}, however, reliably understanding the systematics associated with such calculations as one approaches the physical limit must begin with fully controlling the systematics in the much simpler calculation of $g_A$. In fact, one may argue that the overarching goal of accurately connecting nuclear physics to the SM via lattice QCD relies on the ability to demonstrate control over systematics for well-known quantities. As one of the simplest hadronic matrix elements to calculate, as well as its pervasive appearance in calculations of nearly all nuclear quantities, $g_A$ should be a benchmark.

\begin{figure}
\begin{center}
\includegraphics[width=11.5cm]{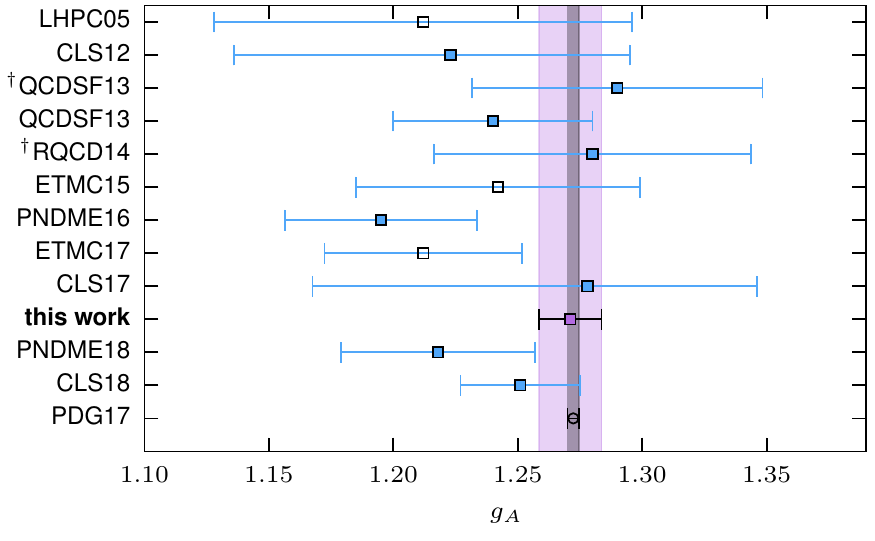}
\end{center}
\vspace{-1cm}
\caption{\label{fig:gA_hist} Selected Lattice QCD results for $g_A$, including the result from Ref.~\cite{Chang:2018gA} described in these proceedings.}
\end{figure}

Through the years, there have been many lattice QCD calculations of $g_A$ appearing from various different collaborations~\cite{Khan:2006de,Lin:2008uz,Capitani:2012gj,Horsley:2013ayv,Bali:2014nma,Abdel-Rehim:2015owa,Alexandrou:2017hac,Edwards:2005ym,Yamazaki:2009zq,Yamazaki:2008py,Bratt:2010jn,Green:2012ud,Yamanaka:2018uud,Ishikawa:2018rew,Bhattacharya:2016zcn,Capitani:2017qpc,Ottnad:2018fri,Liang:2018pis,Gupta:2018wrq,Gupta:2018qil,Berkowitz:2017gql,Chang:2017oll,Chang:2018gA}. Despite the apparent simplicity of such a calculation and the large global effort, predictions of $g_A$ with enough precision to compete with experiment have not been produced. This is to be contrasted with the status of flavor physics in the mesonic sector, for which precision lattice QCD calculations have made a significant impact to our current knowledge~\cite{Aoki:2016frl}. A survey of results for $g_A$ through 2017 is shown in Fig.~\ref{fig:gA_hist}, along with the reported PDG value. 

There are several difficulties which must be overcome in order to produce a precision calculation for $g_A$. One of the most significant differences in lattice calculations involving mesons versus nucleons is that nucleons come with an exponentially poor signal-to-noise problem. This issue is caused by a signal which falls off exponentially with Euclidean time at a faster rate than the Monte Carlo noise, leading to a time dependence of the signal-to-noise ratio which is given by~\cite{Lepage:1989hd}
\beq\label{eq:SN}
S/N \sim e^{-\left(m_N-3/2 m_{\pi} \right) t}  
\eeq
in the limit of large Euclidean time. However, at early Euclidean times the signal is contaminated by contributions from excited states. Thus, in order to minimize systematic errors stemming from excited states, exponentially large computational resources are typically used to clarify the signal at late times. Another way to mitigate this enormous cost is to perform calculations at pion masses that are heavier than those in nature, where the signal-to-noise ratio is exponentially larger, and perform extrapolations to the physical point. For a given set of computational resources, typically some balance must be made between statistical precision and systematic contamination. Unfortunately as calculations move closer to the physical point, whether in Euclidean time or pion mass, the error bars of the calculation grow and may mask systematic effects. 

The work presented here of a new calculation by the CalLat collaboration relies on several advancements to produce a precision of 1\% on the calculation of $g_A$. The first is a new calculation technique based on the Feynman-Hellmann theorem demonstrated in Ref.~\cite{Bouchard:2016heu}. This method, which shares similarities with methods used in Refs.~\cite{Chambers:2014qaa, Chambers:2015bka, Savage:2016kon}, produces, from a single calculation, data points for all possible Euclidean time separations between the nucleon creation and annihilation operators. With access to the full time dependence of the correlation functions, fits to a known functional form for this time dependence may be performed over varying time regions and with different numbers of included excited states, in order to fully understand and eliminate contamination from such excited states. The new technique also allows for multiple operators, each having different overlap onto the ground and excited states, to be produced without the cost of additional calculations. This gives an extra handle for controlling excited states. In addition to lending more confidence in the quoted result and a more reliable systematic uncertainty, the method allows for the use of earlier time separations, in which the signal is exponentially larger, in order to improve the precision of the final result. 

The calculation also utilized a new lattice discretization~\cite{Berkowitz:2017opd}, in which the freely available HISQ ensembles produced by the MILC collaboration~\cite{Bazavov:2012xda} were coupled with propagators calculated using so-called domain-wall fermions (DWF)~\cite{Kaplan:1992bt,Shamir:1993zy,Furman:1994ky}. DWF are preferable for their chiral symmetry properties and the minimization of systematic effects due to the finite lattice spacing, with errors beginning at $\mathcal{O}\left(a^2\right)$, rather than $\mathcal{O}\left(a\right)$ for other types of discretization. Thus, even at a larger lattice spacing a calculation performed using DWF may be effectively closer to the continuum limit than a calculation performed using a different discretization at a smaller lattice spacing. This mixed-action approach balances the cost associated with producing gauge field configurations with the reduction of systematic effects due to discretization. Any additional systematic effects arising from the mixed-action formulation may be estimated or removed using EFT.

The calculation was performed on 16 lattice ensembles, with lattice spacings ranging from $0.09 - 0.15$~fm, pion masses from $130 - 450$~MeV, and volumes between $2.4 - 5.8$~fm. In addition to the MILC ensembles, which range up to $m_{\pi}\sim 310$~MeV, heavier pion mass points were produced to better understand and control the pion mass extrapolations. The pion mass dependence of the calculations is shown in Extended Data Fig. 4 of Ref.~\cite{Chang:2018gA}. 

%\begin{figure}
%\includegraphics[width=0.4\linewidth]{extrap}
%\includegraphics[width=0.4\linewidth]{models}
%\caption{\label{fig:extrap} (left) The model averaged extrapolation of $g_A$ as a function of $\e_\pi = m_{\pi}/\left(4\pi F_{\pi}\right)$. Details of the models used for extrapolation may be found in Ref.~\cite{Chang:2018gA}. (right) Extrapolated results for the model average (top) and for each of the models that went into the average (following six results). Below these are stability tests of the result under the addition of further discretization terms, altering the finite volume extrapolation, doubling all prior widths, cutting data for heavy and light pion masses, cutting data for fine and coarse lattice spacings, and adding additional pion mass dependent terms.}
%\end{figure}

Because the convergence of effective field theory approaches applied to nucleon quantities are not well-understood, various forms of simultaneous extrapolations in the pion mass, lattice spacing, and volumes were performed, based on both effective field theory and simple Taylor expansions. A Bayesian model average was used for the final result, with a systematic error assigned due to model selection. Various checks were performed to study the stability of the results due to, for example, cuts in the data at both heavy and light pion masses. The results for all of these studies is shown in Extended Data Fig. 4 of Ref.~\cite{Chang:2018gA}.

The final result, given by
\beq
g_A = 1.2711(103)^s(39)^{\chi}(15)^a(19^v)(04)^I(55)^M \ ,
\eeq
includes errors arising from statistics ($s$), pion mass dependence ($\chi$), discretization ($a$), finite volume ($v$), isospin breaking effects ($I$), and model selection ($M$). The total error is found to be dominated by the statistical uncertainty. Thus, utilizing the same methods, a straightforward path toward reducing the uncertainty of the theoretical calculation of $g_A$ has been established, with the potential for a resolution of the neutron lifetime puzzle in the near future.

While the uncertainty of the present calculation is still not sufficient to directly compete with experiment, we can already use it to place constraints on potential BSM physics arising from right-handed currents. An exploration of such effects was explored in Ref.~\cite{Alioli:2017ces}, and the authors of this work provided an updated plot showing constraints on left- and right-handed BSM currents ($\delta V_{ud}$ and $\xi_{ud}$, respectively), which is shown in Extended Data Fig. 4 of Ref.~\cite{Chang:2018gA}. The yellow band in this Figure, which is produced using the CalLat calculation of $g_A$, is seen to provide the strongest constraint on $\xi_{ud}$, as compared to results from collider experiments, pion decays, and super-allowed beta decays.

%\begin{figure}
%\begin{center}
%\includegraphics[width=0.6 \linewidth]{xiud}
%\end{center}
%\vspace{-1cm}
%\caption{\label{fig:xiud} A recent comparison of constraints from low-energy experiments and colliders found comparable constraints on right-handed BSM currents\cite{Alioli:2017ces}. The authors of this work generously provided an update of their Fig.~12 using our determination of $g_A$. The vertical orange band is the constraint on the right handed coupling ($\xi_{ud}$) from our result. The diagonal red band is from pion decays (long direction; $\pi\rightarrow\mu\bar{\nu}$) and superallowed $0^+\rightarrow 0^+$ nuclear decays, which constrain corrections to the axial (left - right) and vector (left + right) BSM currents respectively, while the blue circle arises from collider constraints on $W-$ and Higgs-boson production (WH) at collision energy $\sqrt{S}=14$~TeV.}
%\end{figure}

\section{Contributions to neutrinoless double beta decay\label{sec:0nubb}}
An alternative approach to searching for new physics involves tests of fundamental symmetries by looking for exceedingly rare processes that are not allowed by the SM. One example of such a search with an enormous worldwide experimental effort is $0\nu\beta\beta$, in which experimentalists look for two simultaneous beta decays with no emission of neutrinos. Such a process, which violates conservation of lepton number, is only allowed if neutrinos are Majorana in nature: that is, if the neutrino is its own antiparticle. Such an observation could elucidate a potential source for the baryon number asymmetry in the early Universe that is necessary for the observed excess of matter over antimatter. For example, shortly after the Big Bang, heavy right-handed partner neutrinos could decay to leptons, giving a source of leptogenesis~\cite{Pascoli:2006ci,Davidson:2008bu}. These excess leptons can then be converted to an excess of baryons through known SM processes (sphalerons).

There are several compelling reasons to believe that neutrinos could be Majorana in nature. One is based on an Effective Field Theory (EFT) treatment of the SM. We know that the SM is not complete, and therefore at some high energy scale operators with dimension greater than 4 should become important. The only dimension-5 operator one can write down that is consistent with all the symmetries of the SM (aside from lepton number) is a Majorana mass term for the neutrinos. It is already known that the SM violates lepton number conservation through sphaleron processes, and we should expect that any process that is not forbidden by symmetry should be expressed in nature. The existence of Majorana neutrinos also gives rise to a number of models (so-called seesaw models) which provide a convenient explanation for why the neutrinos are so much lighter than all other SM particles. Assuming neutrinos have both Dirac and Majorana mass terms, which requires the introduction of right-handed partner neutrinos, leads to a mass matrix whose two eigenstates are Majorana neutrinos with masses given by $m_l \sim M_D^2/M_R$ and $m_h \sim M_R$, where $M_D, M_R$ are the Dirac and right-handed Majorana masses, respectively. Given that we haven't observed right-handed neutrinos in nature implies that their masses are heavy, leading to a mass for the light neutrino which is inversely suppressed by this heavy scale. The seesaw argument is also implied by the (EFT) treatment of the SM: if the Majorana mass term is dimension-5, it must be suppressed by one power of the high energy scale at which we can no longer resolve the short-distance physics.

There are several current and planned experimental efforts searching for $0\nu\beta\beta$ around the world. These experiments utilize different nuclei, typically $^{76}Ge$, $^{130}Te$, or $^{136}Xe$. They also employ a variety of different detector technologies and observational strategies in efforts to determine the most promising methods for observing this process. The next generation of experiments is planned to be at the ton scale for source material, hopefully pushing the lifetime bounds to $10^{27}$ to $10^{28}$ years. The Nuclear Science Advisory Committee's 2015 Long Range Plan provided an outline of the expected discovery regions for current and proposed experiments as a function of the lightest neutrino mass (Fig. 5.2 of Ref.~\cite{Geesaman:2015fha}). The projections are given for two possible mass hierarchies: the standard hierarchy, in which the smaller of the known neutrino mass splittings corresponds to the lightest pair, and the opposite scenario, the inverted hierarchy. As shown in this Figure, next generation ton-scale experiments should be able to explore the entire discovery region for the inverted hierarchy, while only a modest portion of the discovery region for the standard hierarchy can be accessed. Unfortunately, based on a global analysis of neutrino observables, including cosmological constraints, oscillation experiments, as well as the already excluded lifetimes from $0\nu\beta\beta$ experiments, it seems that the standard hierarchy is favored at the two-sigma level~\cite{Capozzi:2017ipn}. However, this picture assumes that only the most widely studied mechanism, that of light neutrino exchange, is behind the decay. There are many other potential mechanisms for the decay that may provide a more optimistic picture for the potential of experiments.

For example, the already mentioned seesaw mechanism requires heavy right-handed neutrinos to account for the lightness of the observed neutrinos, and these heavy particles may also drive the decay. Na\"ively, one might expect that the heavy particle mechanism for $0\nu\beta\beta$ would be inversely suppressed by the mass of the heavy particle. However, note that the light neutrino mechanism for $0\nu\beta\beta$ is proportional to the light neutrino mass (due to a helicity flip), which, from the same seesaw argument, is also inversely suppressed by the heavy right-handed neutrino mass. Therefore, which mechanism dominates depends largely on the details of the model under consideration. There are several BSM models that give mechanisms for $0\nu\beta\beta$ that don't involve neutrinos at all, such as R-parity violating supersymmetric theories\footnote{Note that it doesn't matter what particles are involved in the decay, observation of $0\nu\beta\beta$ ensures lepton number is violated and that the neutrinos are Majorana, as any diagram leading to the emission of two electrons with no neutrinos can always be rearranged into a Majorana mass diagram for a neutrino (the so-called ``black box theorem"~\cite{schechter1982neutrinoless,Nieves:1984sn,Takasugi:1984xr,Rosen:1992qa,Hirsch:2006yk}).}. Such processes imply that not only should we include contributions from these models when planning experiments, but also if we can relate current experimental bounds to the underlying model, we could potentially, for example, place bounds on R-parity violation, which is essential for understanding the stability of the lightest superpartner, a dark matter candidate. Thus, the questions we must answer are, how do we relate short-distance, BSM models to experimental signatures in large nuclei, and in turn, how do we relate experimental signatures to the underlying theory. 

QCD governs the strong interactions at all energy scales, from the high-energy BSM scale, to the low-energy nuclear scales of experiments. Matching the BSM model onto scales relevant for experiment requires the non-perturbative methods of lattice QCD. Unfortunately, lattice QCD will never be able to directly calculate $0\nu\beta\beta$ matrix elements in large nuclei due to the enormous computational demands. As discussed in the previous Section, the number of statistical samples needed grows exponentially not only with Euclidean time, but also with the number of nucleons. Further issues include the very large number of quark degrees of freedom within a nucleus and the enormous number of lattice sites necessary for correctly resolving both short and long distance scales, to name a few. Therefore the usual paradigm for applying lattice QCD to nuclear physics is that one should use the lattice for calculating few nucleon quantities, and these results can then be used as input into, for example, \textit{ab initio} methods for medium mass nuclei, which are then finally matched onto many-body techniques. 

\begin{figure}
\begin{center}
\includegraphics[width=0.6 \linewidth]{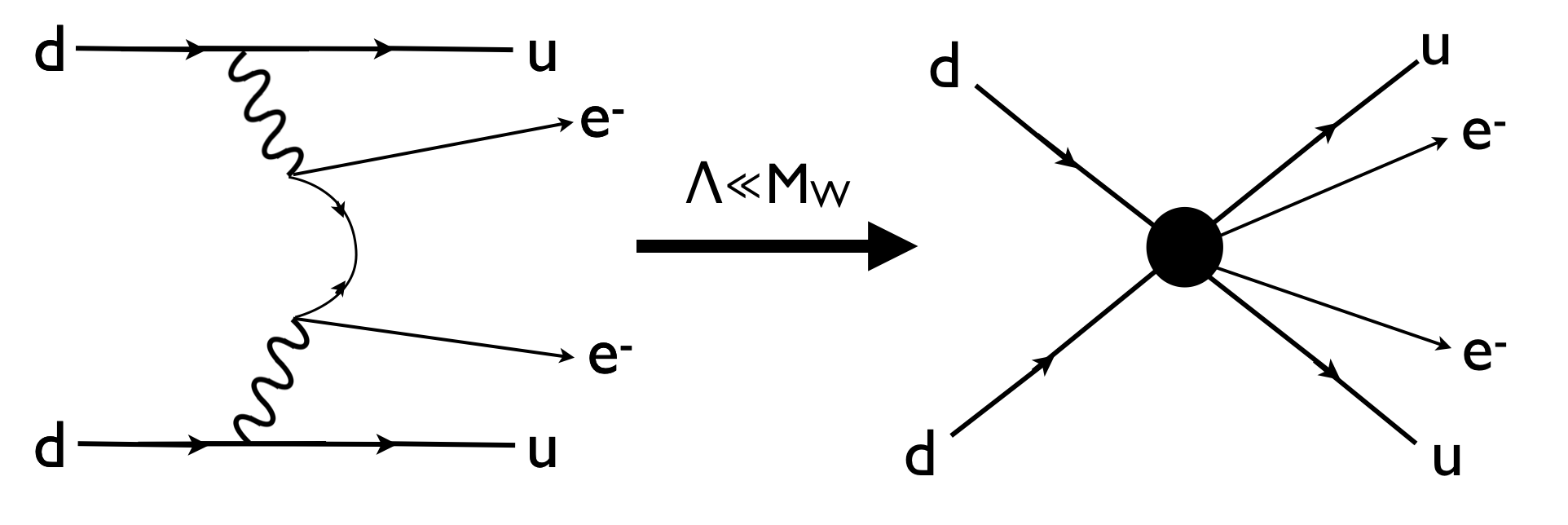}
\end{center}
\vspace{-1cm}
\caption{\label{fig:quarks} Quark-level contributions to $0\nu\beta\beta$. In the left panel, two down quarks are converted to two up quarks plus two electrons via heavy intermediary particles (unlabeled inner lines). At energies much smaller than the electroweak scale, $M_W$, the process appears to occur at a single spacetime point, giving rise to the four-quark diagrams shown in the right panel.}
\end{figure}

The first step in this matching is to determine the matrix elements of two-nucleon systems due to high-energy BSM physics. This matching from the BSM scale to the electroweak scale may be carried out perturbatively. For example, a diagram such as that shown on the left in Fig.~\ref{fig:quarks}, where the unlabeled internal lines belong to some heavy BSM particles, can be represented by the diagram on the right at scales below the electroweak scale. The black circle represents a set of four-quark operators that are consistent with the symmetries of the interaction~\cite{Prezeau:2003xn},
\begin{eqnarray}
\label{eq:Ops}
\mathcal{O}_{1+}^{++} &=& \left(\bar{q}_L \tau^+ \gamma^{\mu}q_L\right)\left[\bar{q}_R \tau^+\gamma_{\mu} q_R \right] \ , \cr
\mathcal{O}_{2\pm}^{++} &=& \left(\bar{q}_R \tau^+ q_L\right)\left[\bar{q}_R \tau^+ q_L \right] \pm \left(\bar{q}_L \tau^+ q_R\right)\left[\bar{q}_L \tau^+ q_R \right] \ , \cr
\mathcal{O}_{3\pm}^{++} &=& \left(\bar{q}_L \tau^+ \gamma^{\mu}q_L\right)\left[\bar{q}_L \tau^+ \gamma_{\mu} q_L \right] \pm \left(\bar{q}_R \tau^+ \gamma^{\mu}q_R\right)\left[\bar{q}_R \tau^+ \gamma_{\mu} q_R \right] \ ,
\end{eqnarray}
where the Takahashi bracket notation $()$ or $[]$ indicates the color indices that are contracted together~\cite{Takahashi:2012}. Note that omitted from this list are vector operators which will be suppressed by the electron mass over a hadronic scale.

\begin{figure}
\begin{center}
\includegraphics[width=0.6 \linewidth]{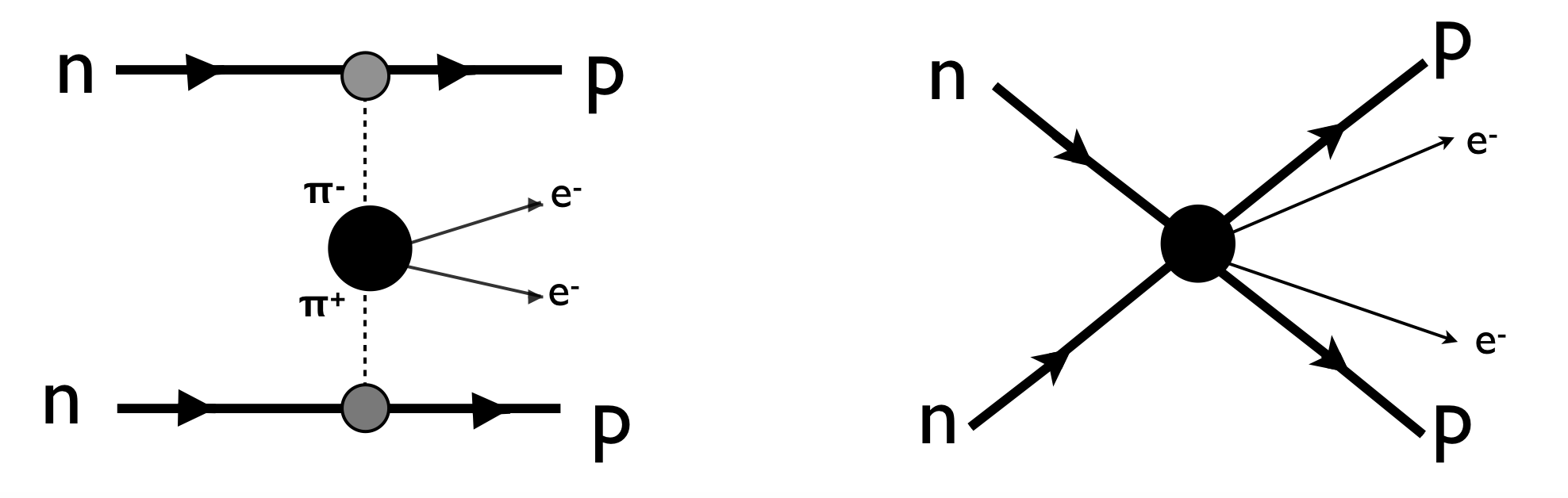}
\end{center}
\vspace{-1cm}
\caption{\label{fig:NN} Two examples of hadronic-level diagrams for short-ranged contributions to $0\nu\beta\beta$. On the left, two neutrons exchange a pion via a known pion interaction vertex (gray circle). The pion undergoes short-ranged $0\nu\beta\beta$ at the black vertex, resulting in two protons and two electrons in the final state. On the right, two neutrons undergo $0\nu\beta\beta$ at a single spacetime point, emerging as two protons and two electrons.}
\end{figure}

To move from the electroweak scale to hadronic scales, there are multiple diagrams one can write down. The leading order contribution, according to Weinberg power counting~\cite{WEINBERG19913,WEINBERG1990288}, is shown on the left side of Fig.~\ref{fig:NN}. Here, the two neutrons exchange a pion, which undergoes $0\nu\beta\beta$, converting it from a $\pi^-$ to a $\pi^+$. At next-to-next-to leading order (NNLO) we have the diagram on the right of this Figure, consisting of two neutrons interacting at a point while undergoing $0\nu\beta\beta$, as well as additional contributions involving pion exchange with a derivative interaction. When calculating the hadronic contributions, one must also consider the so-called color-mixed operators,
\begin{eqnarray}\label{eq:Ops_prime}
\mathcal{O}_{1+}^{'++} &=& \left(\bar{q}_L \tau^+ \gamma^{\mu}q_L\right]\left[\bar{q}_R \tau^+\gamma_{\mu} q_R \right) \ , \cr
\mathcal{O}_{2\pm}^{'++} &=& \left(\bar{q}_L \tau^+ q_L\right]\left[\bar{q}_L \tau^+  q_L \right) \pm \left(\bar{q}_R \tau^+ q_R\right]\left[\bar{q}_R \tau^+ q_R \right) \ ,
\end{eqnarray}
which arise through renormalization from the electroweak scale to the QCD scale\footnote{The corresponding primed operator for $\mathcal{O}_{3\pm}^{++}$ is redundant, which may be shown by performing a Fierz transformation on the resulting operator.}~\cite{Graesser:2016bpz}. 

Considering the leading order calculation on the left side of Fig.~\ref{fig:NN}, there are two types of vertices. The gray circle represents the interaction between a pion and a nucleon, and can be related to $g_A$ through the Goldberger-Trieman relation, Eq.~\ref{eq:GT}. Thus, the only unknowns are the matrix elements for the $\pi^- \to \pi^+$ transition involving the four-quark operators detailed above. Once these matrix elements are determined from lattice QCD, they may be inserted into the EFT diagrams to calculate the $0\nu\beta\beta$ transition rates for two nucleon systems. 

This calculation was performed for the first time in Refs.~\cite{Nicholson:2016byl,Nicholson:2018mwc}. This single pion calculation is computationally far easier to perform on the lattice than full two nucleon calculations due to the lack of any signal-to-noise problem associated with pions and of any complications due to the interpretation of two particle systems in a finite volume~\cite{Briceno:2015tza}. There are five matrix elements to be determined, from the positive parity operators of Eqs.~\ref{eq:Ops},~\ref{eq:Ops_prime}. The results of the lattice calculations for these five operators are presented in Fig. 3 of Ref.~\cite{Nicholson:2018mwc} as a function of the pion mass. 

As the renormalization scale is varied below the electroweak scale, there is mixing between the operators in Eq.~\ref{eq:Ops} and their corresponding primed operators in Eq.~\ref{eq:Ops_prime}. These pairs of operators are shown in green ($\mathcal{O}_{1+}^{++}, \mathcal{O}_{1+}^{'++}$) and blue ($\mathcal{O}_{2+}^{++}, \mathcal{O}_{2+}^{'++}$). $\mathcal{O}_{3+}^{++}$ vanishes at leading order, therefore its value is suppressed compared to the other operators, and the expected $m_{\pi}^2$ dependence is observed. The calculation was performed on a subset of the same ensembles used for the calculation of $g_A$ discussed in the previous Section. Three values of the lattice spacing were used for the continuum extrapolation, along with several different volumes to account for finite volume corrections. 

%\begin{figure}
%\includegraphics[width=0.99\columnwidth]{LR}
%\includegraphics[width=0.99\columnwidth]{S}
%\includegraphics[width=0.48\columnwidth]{O_i}
%\includegraphics[width=0.48\columnwidth]{V}
%\caption{\label{fig:O3}
%The interpolation of the various matrix elements: green ($\mathcal{O}_{1+}^{++}, \mathcal{O}_{1+}^{'++}$), blue ($\mathcal{O}_{2+}^{++}, \mathcal{O}_{2+}^{'++}$). and red ($\mathcal{O}_{3+}^{++}$). In the right panel, a zoomed in version of $O_3$ is displayed.
%The bands represent the 68\% confidence interval of the continuum, infinite volume extrapolated value of the matrix elements.
%The vertical gray band highlights the physical pion mass.
%}
%\end{figure}

Simultaneous extrapolations in the lattice spacing, pion mass, and volume are shown as shaded bands. The final results, given in two different renormalization schemes using non-perturbative renormalization techniques, are given in Table~\ref{tab:O_i}. We find that the values agree at the two sigma level with a recent calculation that utilized $SU(3)$ flavor symmetry and EFT to relate these matrix elements to lattice-determined matrix elements involving kaons~\cite{Cirigliano:2017ymo}.
%%%%%%%%%%%%%%%%%%
%    O_i Table
\begin{table}
\caption{\label{tab:O_i} Resulting matrix elements extrapolated to the physical point, renormalized in RI/SMOM and $\bar{MS}$, both at $\mu=3$~GeV.}
\begin{ruledtabular}
\begin{tabular}{ccc}
& RI/SMOM& $\bar{MS}$ \\%& 1701.01443\\
${O}_i [\textrm{GeV}]^4$& $\mu=3$~GeV& $\mu=3$~GeV\\
\hline
$O_1$            & $-1.91(13)\times10^{-2}$ & $-1.89(13)\times10^{-2}$ \\
%O_4& $-(2.6\pm0.8\pm0.8)\times10^{-2}$ \\
$O_1^\prime$& $-7.22(49)\times10^{-2}$ & $-7.81(54)\times10^{-2}$ \\
%O_5& $-(11\pm2\pm3)\times10^{-2}$ \\
$O_2$           & $-3.68(31)\times10^{-2}$ & $-3.77(32)\times10^{-2}$ \\
%2O_2 & $-(5.4\pm0.6\pm1.0)\times10^{-2}$\\
$O_2^\prime$& $ \phantom{-}1.16(10)\times10^{-2}$ & $ \phantom{-}1.23(11)\times10^{-2}$ \\
%2O_3& $(1.8\pm0.2\pm0.4)\times10^{-2}$\\
$O_3$           & $ \phantom{-}1.85(10)\times10^{-4}$ & $ \phantom{-}1.86(10)\times10^{-4}$ \\
%2O_1& $(2\pm0.2\pm0.4)\times10^{-4}$
\end{tabular}
\end{ruledtabular}
\end{table}
%%%%%%%%%%%%%%%%%%

We have presented the complete lattice calculation necessary for determining $0\nu\beta\beta$ processes involving short-ranged operators within the leading order $\pi^- \to \pi^+$ exchange diagrams. Work remains to be done to faithfully use these results for determining $0\nu\beta\beta$ decay rates within experimentally relevant nuclei. For example, it is known that the Weinberg power counting scheme can be poorly convergent, particularly within the isotriplet $s$-wave channel relevant for this process. This may require the calculation of the full two-nucleon contact diagrams (Fig.~\ref{fig:NN}, right), which can be promoted to a leading order contribution~\cite{Cirigliano:2018hja,Cirigliano:2018yza}. Furthermore, for BSM models in which the $O_1, O_2$ operators vanish, such as left-right symmetric models with no mixing between left- and right-handed $W$ bosons, only NNLO diagrams contribute, thus the full NNLO calculation would need to be performed. Note that the $O_3$ $\pi^- \to \pi^+$ calculation presented here is one element of the NNLO calculation. Finally, matching onto appropriate many-body theories will allow us to propagate these results to large nuclei.

\section{Summary\label{sec:summary}}

We have presented two complete lattice QCD calculations at the physical point for quantities relevant to experimental searches for BSM physics and tests of fundamental symmetries of the SM. All systematic errors in each calculation are controlled and accounted for. We have calculated the axial coupling of the nucleon, $g_A$, with 1\% uncertainty and the promise of a straightforward path to the sub-percent precision necessary for discriminating between discrepant experimental measurements of the neutron lifetime. We have also presented calculations of matrix elements necessary for determining contributions of short-ranged operators to $0\nu\beta\beta$. 

These calculations highlight the present utility of lattice QCD as a tool for complementing experimental progress in the search for new physics. We have entered a precision era, in which all systematics of lattice calculations may be controlled and the results directly compared to nature, giving us a faithful line of communication between theory and experiment.

% ------------------------------------------------------------------
\vspace{12 pt}
\noindent {\sc Acknowledgments:}~
We would like to thank Kostas Orginos for his contributions to Refs.~\cite{Chang:2018uxx,Berkowitz:2018gqe} that were presented in this talk.
We thank the MILC Collaboration for providing their HISQ configurations\cite{Bazavov:2012xda} without restriction.
Numerical calculations were performed with the \texttt{Chroma} software suite~\cite{Edwards:2004sx} with \texttt{QUDA} inverters~\cite{Clark:2009wm,Babich:2011np} on Surface and Vulcan at LLNL, supported by the LLNL Multiprogrammatic and Institutional Computing program through a Tier 1 Grand Challenge award, and on Titan, a resource of the Oak Ridge Leadership Computing Facility at the Oak Ridge National Laboratory, which is supported by the Office of Science of the U.S. Department of Energy under Contract No. DE-AC05-00OR22725, through a 2016 INCITE award.
The calculations were efficiently interleaved with those in Ref.~\cite{Berkowitz:2017gql,Chang:2017oll,Chang:2018gA} using \texttt{METAQ}~\cite{Berkowitz:2017vcp}.

This work was supported by the NVIDIA Corporation (MAC), the DFG and the NSFC through funds provided to the Sino-German CRC 110 ``Symmetries and the Emergence of Structure in QCD'' (EB), a RIKEN SPDR fellowship (ER), the Leverhulme Trust (NG), the U.S. Department of Energy, Office of Science: Office of Nuclear Physics (EB, DAB, CCC, TK, HMC, AN, ER, BJ, PV, AWL,CM); Office of Advanced Scientific Computing (EB, BJ, TK, AWL); Nuclear Physics Double Beta Decay Topical Collaboration (DAB, HMC, AWL, AN); and the DOE Early Career Award Program (DAB, CCC, HMC, AWL) and the LLNL Livermore Graduate Scholar Program (DAB).
This work was performed under the auspices of the U.S. Department of Energy by LLNL under Contract No. DE-AC52-07NA27344 (EB, ER, PV), and by LBNL under Contract No. DE-AC02-05CH11231, under which the Regents of the University of California manage and operate LBNL.
This research was supported in part by the National Science Foundation under Grant No. NSF PHY15-15738 (BCT) and NSF PHY-1748958, and parts of this work were completed at the program ``Frontiers in Nuclear Physics'' (NUCLEAR16).
Part of this work was performed at the Kavli Institute for Theoretical Physics supported by NSF Grant No. PHY-1748958.

%\raggedright

\bibliography{cipanp}

\end{document}